\documentclass[12pt]{article}
\usepackage[T1]{fontenc}
\usepackage[utf8]{inputenc}
\usepackage{lmodern}
\usepackage{amsmath}
\usepackage{amssymb}
\usepackage{graphicx}
\usepackage{caption}
\usepackage{subcaption}
\usepackage{multirow}
\usepackage{float}
\usepackage{mathcomp}
\usepackage{indentfirst}
\title{Interacting dyon ensemble and confinement by particle mesh Ewald's method}
\author{Motahareh Kiamari$^{1}$, Sedigheh Deldar$^{1}$ \\
$^1$Department of Physics, University of Tehran,\\
\small P.O. Box 14395/547, Tehran 1439955961,
Iran.}
\date{}
\begin{document}
\maketitle

\begin{abstract}
The free energy of a static quark-antiquark pair is obtained in an interacting dyon ensemble near the deconfinement temperature. Comparing the results with the noninteracting case, we observe that the string tension between the quark-antiquark pair increases for the interacting ensemble. As a result, the confinement temperature decreases. 
\end{abstract}

\section{Introduction}

Calorons - as one of the candidates of QCD vacuum structure - were first introduced in a set of papers by Diakonov and Petrov \cite{1}\cite{2}\cite{3} to describe quark confinement. They studied the noninteracting ensemble of calorons to calculate the Polyakov loop correlator and obtained the free energy of static quark-antiquark pairs. They also found the temperature of the confinement-deconfinement phase transition by considering the Polyakov loop as an order parameter. However, since the interaction of calorons inside the region of their cores are complex, the interacting ensemble of calorons remained unstudied. This is basically because the core structure of calorons are nonlinear and they are neutral objects without any interactions outside the core. On the other hand, Bruckmann \textit{et al}. \cite{4} showed that the metric introduced by Diakonov and Petrov \cite{1} for the noninteracting calorons, is only positive definite for dyons of different charges or for dyons of the same charge at separations larger than the $ \frac{2}{\pi T} $ in the SU(2) gauge group. They \cite{5} used a numerical method called Ewald's method \cite{6} to solve the problem. For interacting ensembles, they suggested the particle mesh Ewald's (PME) method which is more efficient from the point of view of running time cost. The main idea of Ewald's method is to split the interaction into a converging short-range term and a smooth long-range term which is convergent in the Fourier space.

Applying this method, Bruckmann \textit{et al}. \cite{5} obtained the free energy of static quark-antiquark pairs versus their separations by calculating the Polyakov loop correlator of a noninteracting dyon gas. They also showed that the finite-size volume effects were under control in their calculations.
 
In Ref. \cite{12}, we applied the particle mesh Ewald's method to noninteracting ensemble of dyons and showed that this method also works very well for calculating the free energy between a static quark-antiquark pair. We got a linear rising potential with a well-behaved string tension decreases with increasing temperature.

In this paper we apply the PME method to an interacting dyon ensemble and compare the results with the noninteracting case. For a noninteracting dyon ensemble, the Polyakov loop correlator is calculated by the temporal gauge field of each dyon  whereas the dyons themselves do not interact with each other. For the interacting case, we consider some Coulomb-like interaction between dyons. Our results show that the free energy of the static quark-antiquark pair is also linear for the interacting dyon ensemble, as expected. Comparing the results obtained from the noninteracting and interacting ensembles, we show that by adding the dyonic interactions, the string tension between the quark-antiquark pair increases and therefore the confinement temperature decreases.

The paper is organized as follows. In Sec. \ref{sec:dyon}, some features of dyons are introduced and the Polyakov loop correlator and the action are derived. Ewald's method and the particle mesh Ewald's method are described briefly in Sec. \ref{sec:em}. The setup of our simulations and the numerical results are presented in Sec. \ref{sec:results}. The conclusion and discussions are given in Sec. \ref{sec:Conclusions}.  

\section{Interacting Dyon ensemble for SU(2) Yang-Mills theory}
\label{sec:dyon}

KvBLL calorons - found by Kraan and van Baal \cite{7}, as well as Lee and Lu \cite{8} - are the periodic solutions of the finite-temperature Yang-Mills theory. These neutral objects consist of $N$ dyons in the $SU(N)$ group and have non-Abelian and nonlinear cores which makes it difficult to study their interactions. Dyons are basically non-Abelian objects, but in the far-field limit, they can be considered as $U(1)$ objects with Coulombic electric and magnetic fields. Using the Abelian temporal gauge field in the third direction of color space in $SU(2)$,
\begin{equation}
A_{4}\rightarrow 2\pi\omega T  \sigma_{3},
\label{A4}
\end{equation}
\begin{equation}
\pm B=E\rightarrow \frac{q}{r^{2}} \sigma_{3}, 
\label{ebfield}
\end{equation}  
where \textit{T} is the temperature, $\sigma _{3}$ is the third Pauli matrix and $\omega $ is the holonomy which specifies the confinement and deconfinement phases. The Polyakov loop, 
\begin{equation}
P(\textbf{r})=\frac{1}{2}Tr\left(\exp\left(i\int_{0}^{1/T} dx_{4}A_{4}\left(x_{4},\textbf{r}\right) \right) \right) 
\label{polyakovloop}    
\end{equation}  
is related to the holonomy in the far-field limit,
\begin{equation}
P(\textbf{r})\rightarrow \frac{1}{2}Tr\left(\exp\left(2\pi i\omega\sigma _{3}\right)\right)=\cos\left(2\pi\omega\right).  
\end{equation}
The free energy versus Polyakov loop is defined as
\begin{align}
F_{\bar{Q}Q}(d)=-T\ln \left\langle P(\textbf{r})P^{\dag} (\textbf{r}')\right\rangle , d\equiv \lvert \textbf{r}-\textbf{r}' \rvert ,
\label{free-energy}
\end{align}
where \textit{d} is the distance between a quark located in $\textbf{r}$ and an antiquark in $\textbf{r}'$. Hence, for maximally nontrivial holonomy, where $ \omega=\frac{1}{4} $, the system is in the confinement phase and $ P(\textbf{r})\rightarrow 0 $. For trivial holonomy, the system is in the deconfinement phase and $ P(\textbf{r})\rightarrow \pm 1 $.

To find the Polyakov loop of Eq. (\ref{polyakovloop}), $A_{4}$ of the dyon ensemble has to be found. The long-range gauge fields of a dyon are Coulombic and Abelian in the third direction of color space,
\begin{align}
a_{4}\left(\textbf{r};q\right)=\frac{q}{r},  a_{1}\left(\textbf{r};q\right)=-\frac{qy}{r\left(r-z\right)},  a_{2}\left(\textbf{r};q\right)=+\frac{qx}{r\left(r-z\right)},  a_{3}\left(\textbf{r};q\right)=0.
\label{dyonpotential}
\end{align}
There are two self-dual dyons in $SU(2)$, with electric and magnetic charges equal to $(+1,+1)$ and $(-1,-1)$ corresponding to the plus sign in Eq. (\ref{ebfield}) and two anti-self-dual dyons with charges $(+1,-1)$ and $(-1,+1)$ corresponding to the minus sign in Eq. (\ref{ebfield}). Since we study self-dual dyons and their electric and magnetic charges are equal, these dyons can be considered as objects with one charge $q=\pm 1$.

Using $a_{4}$ of Eq. (\ref{dyonpotential}), the Polyakov loop of the dyon ensemble in confinement phase is obtained from Eq. (\ref{polyakovloop}),
\begin{align}
P(\textbf{r})=\cos\left(2\pi\omega+\frac{1}{2T}\Phi(\textbf{r})\right),  P(\textbf{r})| _{\omega=1/4 } =-\sin\left(\frac{1}{2T}\Phi(\textbf{r})\right),  
\end{align}
\begin{equation}
\Phi(\textbf{r})\equiv \sum_{i=1}^{2K}\frac{q_{i}}{\lvert \textbf{r}-\textbf{r}_{i}\rvert }.  
\label{phi}
\end{equation}       
Keeping in mind that the original system we study is the ensemble of $K$ calorons, we consider the neutral system of $2K$ dyons: $K$ dyons with charge $q=+1$ and $K$ dyons with charge $q=-1$.

To obtain the free energy of Eq. (\ref{free-energy}), the Polyakov loop correlator should be computed. The expectation value of an observable \textit{O},
\begin{equation}
\langle O \rangle=\frac{1}{Z} \int \left( \prod_{k=1}^{n_{D}} d^{3}r_{k} \right) O\left( \left\lbrace \textbf{r}_{k}\right\rbrace \right) \exp \left[S\left( \left\lbrace \textbf{r}_{k}\right\rbrace \right)\right]  
\label{evofO}
\end{equation}
where \textit{Z} is the partition function, $n_{D}$ is the number of dyons in the system, 
\begin{equation}
Z=\int \left( \prod_{k=1}^{n_{D}} d^{3}r_{k} \right) \exp \left[S\left( \left\lbrace \textbf{r}_{k}\right\rbrace \right)\right].  
\label{partitionfun}
\end{equation}
and \textit{S} is the effective action of the ensemble. For noninteracting dyon gas the effective action is constant for all simulations. For the interacting ensemble, the integration measure should be rewritten as
\begin{equation}
\left(\prod _{k=1}^{n_{D}}d^{3}r_{k}\right)\det(G), 
\label{measure}  
\end{equation} 
where \textit{G} is the moduli space metric. This metric is exactly known for two dyons with different charges or a caloron \cite{7}, but for two dyons with the same charge the metric is approximate \cite{1}. Thus, the moduli-space metric for the two-body interaction is
\begin{equation}
G_{(i,j)}=
           \begin{pmatrix}
              2\pi -\frac{2q_{i}q_{j}}{T{\lvert \textbf{r}_{i}-\textbf{r}_{j}\rvert }}& \frac{2q_{i}q_{j}}{T{\lvert \textbf{r}_{i}-\textbf{r}_{j}\rvert }} \\
              \frac{2q_{i}q_{j}}{T{\lvert \textbf{r}_{i}-\textbf{r}_{j}\rvert }} & 2\pi -\frac{2q_{i}q_{j}}{T{\lvert \textbf{r}_{i}-\textbf{r}_{j}\rvert }},
            \end{pmatrix}
            \label{metric}
\end{equation}
with the eigenvalues,
\begin{align}
\lambda _{1}=2\pi, \lambda _{2}=2\pi -\frac{4q_{i}q_{j}}{T{\lvert \textbf{r}_{i}-\textbf{r}_{j}\rvert }}. 
\end{align}
To have a positive-definite metric, the distance between dyons of the same charge should not be less than $ \frac{2q_{i}q_{j}}{\pi T} $. The determinant of the moduli-space metric is
\begin{equation}
\resizebox{0.99\textwidth}{!}{$\prod _{(i,j)} det(G_{(i,j)}) = \prod _{(i,j)} 4\pi^{2} \left( 1- \frac{2q_{i}q_{j}}{\pi T \lvert \textbf{r}_{i}-\textbf{r}_{j}\rvert} \right) = (4\pi^{2})^{n_{D}^{2}} \exp \left[ \sum _{(i,j)} \ln \left( 1- \frac{2q_{i}q_{j}}{\pi T \lvert \textbf{r}_{i}-\textbf{r}_{j}\rvert} \right) \right]. $}
\end{equation}
Now one can rewrite the expectation value (\ref{evofO}) and the partition function (\ref{partitionfun}),
\begin{equation}
\langle O \rangle=\frac{1}{Z} \int \left( \prod_{k=1}^{n_{D}} d^{3}r_{k} \right) O\left( \left\lbrace \textbf{r}_{k}\right\rbrace \right) \exp \left[S_{eff}\left( \left\lbrace \textbf{r}_{k}\right\rbrace \right)\right]  
\end{equation}
\begin{equation}
Z=\int \left( \prod_{k=1}^{n_{D}} d^{3}r_{k} \right) \exp \left[S_{eff}\left( \left\lbrace \textbf{r}_{k}\right\rbrace \right)\right].  
\end{equation}
where the effective action is,
\begin{equation}
S_{eff}\left( \left\lbrace \textbf{r}_{k}\right\rbrace \right)=\frac{1}{2}\sum _{i=1}^{n_{D}}\sum _{j=1,j\neq i}^{n_{D}} \ln \left(1 -\frac{2q_{i}q_{j}}{\pi T{\lvert \textbf{r}_{i}-\textbf{r}_{j}\rvert }} \right). 
\label{action} 
\end{equation}
\\To include the contribution of anti-self-dual dyons, one should modify the metric of Eq. (\ref{metric}) to a (4\texttimes4) matrix \cite{1}. The diagonal (2\texttimes2) blocks of the new metric describe the same-duality dyons, while the off-diagonal (2\texttimes2) blocks represent the interactions of different-duality dyons. Thus, we should calculate the determinant of this metric with the nonzero off-diagonal elements. All terms in the modified metric are Coulombic and we should apply all steps of Ewald's method to the anti-self-dual dyons, as well. This modification makes the calculations very difficult and cumbersome. However, in Ref. \cite{1} Diakonov showed that adding anti-self-dual dyons only changes the string tension to $\sqrt{2}$ of its value when we do not use them and the physics of the quark-antiquark potential does not change. Therefore, we trust Diakonov's calculations and study the ensemble of $K$ calorons as he did. Our main goal - which is to study the linearity of the free energy and to observe the increasing of the string tension due to the dyonic interactions - will not be affected.

In the next section we calculate the Polyakov correlator with Ewald's method using the partition function and the action we obtained in this section.

\section{Ewald's method}
\label{sec:em}

The first step is applying Ewald's method is to mimic the space with a basic cell called a super cell, and copy it in all three directions and put the particles in the super cell. The copies contain the copies of the particles. This is how the periodic boundary condition is applied. Therefore we put $n_{D}$ dyons randomly in the super cell. The second and main step is to split the long-range term $\frac{1}{r}$ into an exponentially short-range part and a smooth long-range part,
\begin{equation}
\Phi(\textbf{r}) =\Phi ^{\texttt{short}}(\textbf{r})+\Phi ^{\texttt{long}}(\textbf{r}),
\end{equation} 
\begin{equation}
\Phi ^{S}(\textbf{r})\equiv \sum_{\textbf{n}\in \mathbb{Z} ^{3} } \sum_{j=1}^{n_{D}}\left( 1-\texttt{erf}\left( \frac{\lvert \textbf{r}-\textbf{r}_{j} -\textbf{n}L\rvert}{\sqrt{2}\lambda}\right) \right) \frac{q_{j}}{\lvert \textbf{r}-\textbf{r}_{j} -\textbf{n}L \rvert },
\label{short}
\end{equation}
\begin{equation}
\Phi ^{L}(\textbf{r})\equiv \sum_{\textbf{n}\in \mathbb{Z} ^{3} } \sum_{j=1}^{n_{D}}\texttt{erf}\left( \frac{\lvert \textbf{r}-\textbf{r}_{j} -\textbf{n}L\rvert}{\sqrt{2}\lambda}\right) \frac{q_{j}}{\lvert \textbf{r}-\textbf{r}_{j} - \textbf{n}L\rvert },
\label{long}
\end{equation}
where $ \lambda $ is an arbitrary parameter and \texttt{erf} is the error function. The vector \textbf{n} specifies the copies of the super cell and $L^{3}$ is the spatial volume of the super cell. $\Phi ^{S}$ is convergent for a finite cutoff but $\Phi ^{L}$ is a divergent smooth function. Thus, its Fourier transformation is convergent for finite cutoff,
\begin{align}
\Phi ^{L}(\textbf{r})=\frac{4\pi}{L^{3}}\sum_{\textbf{n}\in \mathbb{Z} ^{3} \setminus  \vec{0}} \frac{e^{-\lambda ^{2}\textbf{k}(\textbf{n})^{2}/2}}{\textbf{k}(\textbf{n})^{2}}Re\left( \sum_{j=1}^{n_{D}} q_{j}e^{+i\textbf{k}(\textbf{n})\textbf{r}}e^{-i\textbf{k}\textbf{}(\textbf{n})\textbf{r}_{j}}\right) , \textbf{k}(\textbf{n})\equiv \frac{2\pi }{L}\textbf{n} .
\label{ploop}
\end{align} 
where 
\begin{equation}
S(k)= \sum_{j=1}^{n_{D}} q_{j}e^{-i\textbf{k}\textbf{}(\textbf{n})\textbf{r}_{j}} 
\label{sfactor}
\end{equation}
is called the structure factor. To reduce the operating costs, one needs $\Phi ^{S}(\textbf{r}) $ to be convergent in the original super cell. However, the arbitrary parameter $\lambda $ should be chosen such that $\Phi ^{S}(\textbf{r}) $ converges in a sphere with a maximum radius $r_{max}<L/2$ within an appropriate error \cite{9}. The center of the sphere is located at position \textbf{r}. Consider $ J(\textbf{r})$ which indicates all dyons and copies of them in this sphere,
\begin{equation}
\Phi ^{S}(\textbf{r})\equiv \sum_{j\in J(\textbf{r})} \texttt{erfc}\left( \frac{\lvert \textbf{r}-\textbf{r}_{j} \rvert}{\sqrt{2}\lambda}\right) \frac{q_{j}}{\lvert \textbf{r}-\textbf{r}_{j} \rvert },
\label{short2}
\end{equation}
where $ \texttt{erfc}(x) = 1-\texttt{erf}(x)$. As mentioned before, the long-range part $\Phi ^{L}(\textbf{r})$ converges in Fourier space. Hence, one can consider a sphere with radius $ n_{max}$, in which $\Phi ^{L}(\textbf{r})$ has to converge \cite{9}.

For large dyon separations $r_{ij}$, the action in Eq. (\ref{action}) can be expanded in powers of $ \frac{1}{r} $,
\begin{equation}
S^{N}=\frac{1}{2} \sum_{i\neq j}\left(-\frac{2q_{i}q_{j}}{\pi Tr_{ij}} -\frac{2\left( q_{i}q_{j}\right) ^{2}}{\left( \pi Tr_{ij}\right) ^{2}} - \frac{8\left( q_{i}q_{j}\right) ^{3}}{3\left( \pi Tr_{ij}\right) ^{3}} + O\left( \frac{1}{r_{ij}^{4}}\right) \right),
\label{expansion} 
\end{equation} 
where $ r_{ij}=\lvert \textbf{r}_{i}-\textbf{r}_{j} \rvert  $ and the superscript $N$ denotes the nonperiodic summation of the action. The action of Eq. (\ref{expansion}) has the $\frac{1}{r^{p}}$, $ p\in \mathbb{R}$ terms, so to apply Ewald's method to calculate these terms we should generalize the above procedure to the $\frac{1}{r^{p}}$ terms. With the definition of the Euler gamma function and the Fourier integral expansion of the three-dimensional Gaussian distribution, one can obtain the $ \frac{1}{r^{p}} $ term, 
\begin{equation}
\frac{1}{r^{p}}=\frac{\pi ^{3/2}}{\left(\sqrt{2}\lambda\right)^{p-3}} \int d^{3}\textbf{u}f_{p}\left(\sqrt{2}\lambda\pi \lvert \textbf{u}\rvert \right) \exp\left(-2i\pi\textbf{u}.\textbf{r}\right)+\frac{g_{p}\left(r/\sqrt{2}\lambda\right)}{r^{p}},
\label{rinv}  
\end{equation} 
where
\begin{equation}
g_{p}(x)=\frac{2}{\Gamma\left(p/2\right)}\int_{x}^{\infty } s^{p-1}\exp (-s^{2})ds 
\label{gp}
\end{equation}
\begin{equation}
f_{p}(x)=\frac{2x^{p-3}}{\Gamma\left(p/2\right)}\int_{x}^{\infty } s^{2-p}\exp (-s^{2})ds.
\label{fp} 
\end{equation}
The first and the second terms of Eq. (\ref{rinv}) express the long-range part and the short-range part, respectively. This is because $ \lim _{x\rightarrow \infty} g_{p}(x) = 0 $ while $ \lim _{x\rightarrow \infty} f_{p}(x) \neq  0 $. Using Eq. (\ref{rinv}) for each term of Eq. (\ref{expansion}) and using periodic boundary conditions, one can split the terms of the action into the short-range term, long-range term, and self-energy term,
\begin{equation}
S_{p}^{P}=\sum_{l=1}^{p}\left(S^{S}_{(l)}+S^{L}_{(l)}-S^{self}_{(l)}\right), 
\label{periodicS} 
\end{equation}
where the superscript \textit{P} denotes the periodic summation of the action that consists of the copies of dyons in copies of the super cell. We should modify the formula in Ref. \cite{9} since the charges of the dyons in that reference are $\pm 1$ and therefore the multiplication of charges in the numerators of equations like (\ref{expansion}) is equal to 1 for the even power of the charges. As a result, the only odd power of the charges is 1. While in our case, we are dealing with the interpolated charges with different values which depend on the positions of the randomly located dyons for each configuration. The interpolated charges are introduced in the next section. We should also add the self-energy part to the action. This is because the self-energy is a function of the power of the charges [Eq. (\ref{self-s})]. These terms have different and important values for our dyons, while for dyons with $\pm 1$ charges the self-energy terms are constant for the simulations with a fixed number of dyons and thus they do not affect the correlation function of Eq. (\ref{evofO}):
\begin{equation}
S^{S}_{(l)} =  c(l)\frac{1}{2} \sum_{\textbf{n}\in \mathbb{Z} ^{3} }\sum_{i\neq j} \frac{q_{i}^{l}q_{j}^{l}}{\lvert \textbf{r}_{i}-\textbf{r}_{j} -\textbf{n}L\rvert^{l}} g_{l} \left( \frac{\lvert \textbf{r}_{i}-\textbf{r}_{j} -\textbf{n}L\rvert}{\sqrt{2}\lambda } \right),
\label{SS}
\end{equation}
and $c(l)$ is the coefficient of the $l$th term in Eq. (\ref{expansion}),
\begin{equation}
S^{L}_{(l)} =  c(l) \frac{\pi ^{3/2}}{2V (\sqrt{2}\lambda )^{l-3}} \sum_{\textbf{k}_{sym}} f_{l} \left( \frac{\lambda k}{\sqrt{2}} \right) \left( 2\lvert S(\textbf{k},l)\rvert^{2} \right).
\label{LS}
\end{equation}
$ S(\textbf{k},l)=\sum_{i=1}^{n_D}q^{l}_{i}e^{-i\textbf{k}.\textbf{r}_{i}} $ and \textbf{k} is symmetric with respect to $\textbf{k}=0$, and the summation on $ \textbf{n}$ is done by the term $ \exp\left(-2i\pi\textbf{u}.\textbf{n}L \right) $ of Eq. (\ref{rinv}),
\[ \sum _{\textbf{n}} \exp(-2\pi i \textbf{u}\textbf{n}L ) = \frac{1}{V} \sum _{m}^{\infty} \delta \left( \textbf{u} - \frac{\textbf{m}}{L} \right),\]
since \textbf{u} is the reciprocal vector, and the integral on \textbf{u} in Eq. (\ref{rinv}) changes all \textbf{u} to $ \frac{m}{L}$, where $ \textbf{k} = 2\pi \frac{\textbf{m}}{L}$. The self-energy part of the short-range term can be canceled by omitting the $ i=j $ term, but the self-energy part of the long-range term should be separated. This term is the long-range part of the energy when $ \textbf{r}_{j}-\textbf{r}_{i}\rightarrow 0 $. In general, this term can be obtained by subtracting the short-range part in Eq. (\ref{gp}) from the total term $ \frac{1}{r^{p}} $,
\begin{equation}
\lim_{r\rightarrow 0} \left( \frac{1}{r^{p}}-\frac{g_{p}\left( r/\sqrt{2}\lambda\right) }{r^{p}}\right) = \frac{2\left( \sqrt{2}\lambda\right)^{p} }{p \Gamma \left( p/2\right) }, 
\label{self} 
\end{equation}
which gives
\begin{equation}
S^{self}_{(l)} = \frac{2(1/\sqrt{2}\lambda)^{p}}{p\Gamma (p/2)} c(l) \sum_{i=1}^{n_{D}} q_{i}^{l}.
\label{self-s}
\end{equation}
Now, for $l=1,2,3$,
\begin{equation}
S^{S}_{(1)}=-\frac{1}{\pi}\sum_{i=1}^{n_{D}} \sum_{j\in J(\textbf{r}_{i})} \frac{q_{i}q_{j}}{Tr_{ij}} \texttt{erfc}\left(\frac{r_{ij}}{\sqrt{2}\lambda } \right),
\label{S1S}  
\end{equation}
\begin{equation}
S^{S}_{(2)}=-\frac{1}{\pi^{2}}\sum_{i=1}^{n_{D}} \sum_{j\in J(\textbf{r}_{i})} \frac{q^{2}_{i}q^{2}_{j} }{T^{2}r_{ij}^{2}}\exp \left(-\frac{r_{ij}^{2}}{2\lambda ^{2}}\right), 
\label{S2S} 
\end{equation}
\begin{equation}
S^{S}_{(3)}=-\frac{4}{3\pi^{3}}\sum_{i=1}^{n_{D}} \sum_{j\in J(\textbf{r}_{i})} q^{3}_{i}q^{3}_{j}\left( \frac{\texttt{erfc}\left( \frac{r_{ij}}{\sqrt{2}\lambda  }\right)}{T^{3}r_{ij}^{3}}+ \sqrt{\frac{2}{\pi }} \frac{\exp \left(-\frac{r_{ij}^{2}}{2\lambda ^{2}}\right)}{T^{3}\lambda r_{ij}^{2}}\right) , 
\label{S3S} 
\end{equation}
\begin{equation}
S^{L}_{(1)}=-\frac{8}{TV}\sum_{\textbf{k}sym} \lvert S(\textbf{k},1)\rvert^{2} \frac{\exp \left( -\frac{\lambda ^{2}\textbf{k}^{2}}{2} \right)}{\textbf{k}^{2}}, 
\label{S1L} 
\end{equation}
\begin{equation}
S^{L}_{(2)}=-\frac{4}{T^{2}V}\sum_{\textbf{k}sym} \lvert S(\textbf{k},2)\rvert^{2} \frac{\texttt{erfc} \left( \frac{\lambda k}{\sqrt{2}} \right)}{k}, 
\end{equation}
\begin{equation}
S^{L}_{(3)}=-\frac{16}{3\pi ^{2}T^{3}V}\sum_{\textbf{k}sym} \lvert S(\textbf{k},3)\rvert^{2} \left(-\texttt{Ei}\left(-\frac{k^{2}\lambda ^{2} }{2}\right)\right), 
\label{S3L} 
\end{equation}
where Ei is the exponential integral $ \texttt{Ei}(x)=-\int _{-x}^{\infty }\frac{e^{-t}}{t}dt $, and
\begin{equation}
S^{self}_{(1)}=\frac{-2}{\sqrt{2}\lambda \pi ^{3/2}}\sum^{n_{D}} _{i=1} q^{2}_{i},
\label{S1self} 
\end{equation}
\begin{equation}
S^{self}_{(2)}=\frac{-1}{2\lambda^{2} \pi ^{2}}\sum^{n_{D}} _{i=1} q^{4}_{i},
\label{S2self} 
\end{equation}
\begin{equation}
S^{self}_{(3)}=\frac{-8}{9\sqrt{2}\lambda^{3} \pi ^{7/2}}\sum^{n_{D}} _{i=1} q^{6}_{i}.
\label{S3self} 
\end{equation}
As mentioned before, the expansion in Eq. (\ref{expansion}) is only appropriate for large dyon separations, and thus for small dyon separations a correction term should be added to the periodic action in Eq. (\ref{periodicS}),
\begin{equation}
S=S_{p}^{P}-S_{p}^{Corr}.
\label{Self} 
\end{equation}
To have a continuous action on the boundary of small and large dyon separations, $ r_{Corr} $, we should subtract the expansion of the action in Eq. (\ref{expansion}) from the periodic result in Eq. (\ref{periodicS}) and add S from Eq. (\ref{action}), 
\begin{equation}
S_{p}^{Corr}=\sum_{j=1}^{n_D}\sum_{i\in I(\textbf{r}_{j})} \left[\sum_{l=1}^{p} S_{(l)}^{N}(q_{i}q_{j},r_{ij})-\frac{1}{2}\ln \left( 1-\frac{2q_{i}q_{j}}{\pi Tr_{ij}}\right)  \right],
\label{correct} 
\end{equation}
because for small $ r_{ij} $, $ S^{P} $ and $ S^{N} $ are approximately equal and for large $ r_{ij} $, $ S^{N} $ and the action in Eq. (\ref{action}) are equal \cite{9}. Here, $ I(\textbf{r}_{j}) $ is the set of dyons and their copies and their separations from the \textit{i}th dyon are less than $ r_{Corr} $, and $ S_{(l)}^{N} $ stands for the \textit{l}th-order term of $ S^{N} $ in Eq. (\ref{expansion}). By expanding Eq. (\ref{correct}), $ S_{p}^{Corr} $ for different values of \textit{p} is found,
\begin{equation}
\begin{split}
S^{corr}_{1} =& \frac{1}{2}\sum_{j=1}^{n_D}\sum_{i\in I(\textbf{r}_{j})}\left[ -\frac{2q\left(\textbf{r}_{i} \right)q\left(\textbf{r}_{j} \right) }{\pi Tr_{ij}}-\left(-\frac{2q\left(\textbf{r}_{i} \right)q\left(\textbf{r}_{j} \right) }{\pi Tr_{ij}} -\frac{2q^{2}\left(\textbf{r}_{i} \right)q^{2}\left(\textbf{r}_{j} \right)}{\pi^{2} T^{2}r^{2}_{ij}} + O \left( \frac{1}{r_{ij}^{3}} \right) \right)\right] {}\\
  =& \frac{1}{2}\sum_{j=1}^{n_D}\sum_{i\in I(\textbf{r}_{j})}\frac{2q^{2}\left(\textbf{r}_{i} \right)q^{2}\left(\textbf{r}_{j} \right)}{\pi^{2} T^{2}r^{2}_{ij}} + O \left( \frac{1}{r_{ij}^{3}} \right).
\end{split}
\label{S1} 
\end{equation}
Performing the same procedure,
\begin{equation}
S^{corr}_{2} = \frac{1}{2}\sum_{j=1}^{n_D}\sum_{i\in I(\textbf{r}_{j})}\frac{8q^{3}\left(\textbf{r}_{i} \right)q^{3}\left(\textbf{r}_{j} \right)}{3\pi^{3} T^{3}r^{3}_{ij}} + O \left( \frac{1}{r_{ij}^{4}} \right),
\label{S2}
\end{equation}
\begin{equation}
S^{corr}_{3} = \frac{1}{2}\sum_{j=1}^{n_D}\sum_{i\in I(\textbf{r}_{j})}\frac{4q^{4}\left(\textbf{r}_{i} \right)q^{4}\left(\textbf{r}_{j} \right)}{\pi^{4} T^{4}r^{4}_{ij}} + O \left( \frac{1}{r_{ij}^{5}} \right).
\label{S3}
\end{equation}
Since we approximate the action terms of Eq. (\ref{expansion}) up to order $ O(r^{3}) $, the correction terms up to $ O(r^{4}) $ are good enough.

To summarize this section, we have obtained the following action for an interacting dyonic system:
\begin{equation}
S = \sum _{l=1}^{p} \left(S^{S}_{(l)}+S^{L}_{(l)}-S^{self}_{(l)}\right)-S_{p}^{Corr},
\label{finalS}
\end{equation}
where $S^{S}_{(l)}$, $S^{L}_{(l)}$, and $S^{self}_{(l)}$ were introduced in Eqs. (\ref{S1S})~(\ref{S3self}), respectively. $S_{p}^{Corr}$ in Eq. (\ref{S3}) is added to the action which represents a correction term corresponding to the small dyon separations. We calculated the action for $p=3$ in the above Eq. (\ref{finalS}) and we have discussed that it is a good approximation. 

\subsection{Particle mesh Ewald's method}
\label{subsec:pme}

The main idea of the particle mesh Ewald's method \cite{10} is to grid the super cell in reciprocal space and interpolate the charge of each particle to the nearest neighboring mesh points. This method was first introduced by Hockney and Eastwood \cite{11} within a computer simulation and is more efficient for interacting dyon gas.

Consider $n_{D}$ dyons distributed randomly in a super cell at positions $ \textbf{r}_{1},\textbf{r}_{2},...,\textbf{r}_{n_{D}} $. Each dyon at position $\textbf{r}_{i}$ in real space has fractional coordinates $s_{\alpha i}=\textbf{a}_{\alpha }^{*}.\textbf{r}_{i}$ in reciprocal space. Then, we grid the super cell by the points $K_{i}$ for each direction. The new scaled fractional coordinates $u_{1}$,$u_{2}$,$u_{3}$ are defined as $ u_{\alpha }=K_{\alpha }\textbf{a}_{\alpha }^{*}.\textbf{r} $, $ \alpha=1,2,3 $, and $ 0\leq u_{\alpha } < K_{\alpha } $ due to the periodic boundary condition. Then, the terms of the structure factor of Eq. (\ref{sfactor}) can be rewritten with these new coordinates. \textit{m} is the reciprocal vector, $ \textbf{m}=m_{1}\textbf{a}_{1}^{*}+m_{2}\textbf{a}_{2}^{*}+m_{3}\textbf{a}_{3}^{*} $,
\begin{equation}
\exp \left( -i\textbf{m}.\textbf{r}\right)=\exp \left(-i\frac{m_{1}u_{1}}{K_{1}} \right).\exp \left(-i\frac{m_{2}u_{2}}{K_{2}} \right).\exp \left(-i\frac{m_{3}u_{3}}{K_{3}} \right).
\label{sfterms} 
\end{equation}
In Ref. \cite{10} both piecewise Lagrangian and cardinal B-Spline interpolations were introduced, but the latter was applied to calculate the energy of the molecular system. This is because the coefficients of this interpolation are $n-2$ times continuously differentiable. $n$ is the number of neighbor mesh points used for interpolation, and the authors needed differentiability to calculate the forces between molecules, while the coefficients of piecewise Lagrangian interpolation are only piecewise differentiable. Since we do not need to calculate the force and therefore differentiability, we apply piecewise Lagrangian interpolation. By this interpolation, these exponentials can be approximated for $p> 1$,
\begin{equation}
\exp \left(-i\frac{m_{\alpha}}{K_{\alpha}}u_{\alpha}\right) \approx \sum _{k=-\infty}^{\infty}W_{2p}(u_{\alpha}-k). \exp\left(-i\frac{m_{\alpha}}{K_{\alpha}}k\right),
\label{exp}
\end{equation}
where $W_{2p}(u^{'})=0$ for $|u^{'}| > p$ and for $-p\leq u^{'} \leq p$ the coefficient $W_{2p}(u^{'})$ is
\begin{equation}
W_{2p}(u^{'}) = \frac{\prod _{j=-p,j\neq k^{'}}^{p-1} (u^{'}+j-k^{'})}{\prod _{j=-p,j\neq k^{'}}^{p-1} (j-k^{'})}, k^{'}\leq u^{'}\leq k^{'}+1, k^{'}=-p,-p+1,...,p-1.
\label{W2p}
\end{equation}
The subscript $2p$ is the order of interpolation and specifies the number of mesh points used to interpolate the $\exp(-i mu/K)$ in each direction. These points are $[u]-p+1$, $[u]-p+2$ , ..., $[u]+p$, which are the $2p$ nearest neighbor mesh points to the point $ u $. Using Eq. (\ref{exp}), one can approximate the structure factor in Eq. (\ref{sfactor}),
\begin{equation}
\begin{split}
S(\textbf{m})\approx & \widetilde{S}(\textbf{m})=\sum _{i=1}^{n_{D}}q_{i} \sum _{k_{1}=-\infty}^{\infty}\sum _{k_{2}=-\infty}^{\infty}\sum _{k_{3}=-\infty}^{\infty} W_{2p}(u_{1i}-k_{1})W_{2p}(u_{2i}-k_{2}) {}\\
&.W_{2p}(u_{3i}-k_{3})\exp \left(-i\frac{m_{1}}{K_{1}}k_{1}\right)\exp \left(-i\frac{m_{2}}{K_{2}}k_{2}\right)\exp \left(-i\frac{m_{3}}{K_{3}}k_{3}\right). 
\end{split}
\label{stilda}
\end{equation}
Comparing the new structure factor of Eq. (\ref{stilda}) with the structure factor of Eq. (\ref{sfactor}), the new charges assigned to the mesh points are
\begin{equation}
   \begin{split} 
Q(k_{1},k_{2},k_{3})=  \sum_{i=1}^{n_D}\sum_{n_{1},n_{2}n_{3}}q_{i} & W_{2p}(u_{1i}-k_{1}-n_{1}K_{1})W_{2p}(u_{2i}-k_{2}-n_{2}K_{2}) {}\\
& .W_{2p}(u_{3i}-k_{3}-n_{3}K_{3}).
   \end{split} 
   \label{Q} 
\end{equation}
The new structure factor is
\begin{equation}
S(\textbf{m})\approx \sum_{k_{1}=0 }^{K_{1}-1 }\sum_{k_{2}=0 }^{K_{2}-1 }\sum_{k_{3}=0 }^{K_{3}-1 } Q(k_{1},k_{2},k_{3}) \exp \left[-i\left(\frac{m_{1}k_{1}}{K_{1}}+\frac{m_{2}k_{2}}{K_{2}}+\frac{m_{3}k_{3}}{K_{3}} \right) \right].
\label{newstructurefac} 
\end{equation}
The structure factor of Eq. (\ref{newstructurefac}) describes the new system with new $K_{1}K_{2}K_{3}$ charges $Q(k_{1},k_{2},k_{3})$ introduced in Eq. (\ref{Q}) which are located on mesh points $(k_{1},k_{2},k_{3})$ on a three-dimensional ($3D$) lattice. We use this system instead of the system with $n_{D}$ dyons located randomly on $r_{i}$. Now, we apply the simple Ewald's method to this new system. The advantage of this new system is the constant number of charges, $K_{1}K_{2}K_{3}$, which are the same in all simulations,  in contrast to the number of dyons $n_{D}$ of the original system which are different for each individual simulation.

\section{Simulation results}
\label{sec:results}

As mentioned in the Introduction, studying quark confinement with dyons as the constituents of the QCD vacuum is the main purpose of this article. Using the Polyakov loop correlator of Sec. \ref{sec:dyon}, the free energy of a static quark-antiquark pair is calculated for both non interacting and interacting dyon ensembles. Ewald's method (introduced in Sec. \ref{sec:em}) is applied to the system of the charges obtained with the PME method in Sec. \ref{subsec:pme}, for dyons located randomly on 3D lattice. Before applying the particle mesh Ewald's method, we need to put some dyons randomly in a super cell on the lattice. To make sure that dyons are sitting randomly in the super cell, we also use a Metropolis algorithm to make sure the system is in a stable energy. We do this procedure for each configuration before applying Ewald's method and the dyonic interaction.  

$n_{D}$ dyons are assumed to be located randomly in a super cell in the following procedure:
\\1. Fill the super cell with $N$ dyons with random 3D coordinates.
\\2. Displace one dyon slightly.
\\3. Compute the change of the action due to this displacement, $\Delta S$.
\\4. If $\Delta S < 0 $, accept the new configuration.
\\5. If $\Delta S > 0 $, accept the new configuration with the conditional probability: pick a random number $0<x<1$; if $\exp(-\Delta S) > x $, accept the new configuration; if $\exp(-\Delta S) < x $, reject the new configuration.
\\We should mention that in this procedure we calculate only the part of the action related to the dyon which is displaced. For each configuration, we perform steps 2 to 5 for all $N$ dyons.

 We interpolate the charges of these dyons to the 3D lattice with $K_{i}=16$ as described in Sec. \ref{subsec:pme}. This interpolation leads the system to a new setup with charges located on the mesh points according to Eq. (\ref{Q}). Since the structure factors of these old and new systems are approximately equal,  the two systems are equivalent and we use the new system of interpolated charges instead of the original old system of dyons. We apply Ewald's method to this new system to calculate both the short-range and long-range parts of the Polyakov loop and also the action introduced in Sec. \ref{sec:dyon}, while in Ref. \cite{10} the PME method was only applied to calculate the long-range part of the action. Using this method, we do not have to increase the number of mesh points even for a large number of dyons, since for any number of dyons we can interpolate them to a constant number of mesh points. This saves on operating costs, in contrast to the case where one puts dyons directly on a lattice and increases the lattice points as the number of dyons increases \cite{5}. 
We fix the dyon density $ \rho $ and temperature $ T $ to $ \rho /T^{3} = 1 $ which scales the separations by $ \rho ^{1/3} $ or $T$, as done in Ref. \cite{5}. Various lattice sizes, the number of configurations, the number of dyons and other parameters of our simulations are listed in Table \ref{tab:input}.

\begin{figure}
\captionsetup{font=footnotesize}
\centering
\begin{subfigure}{.45\textwidth}
  \centering
  \includegraphics[width=1\linewidth,height=1\linewidth]{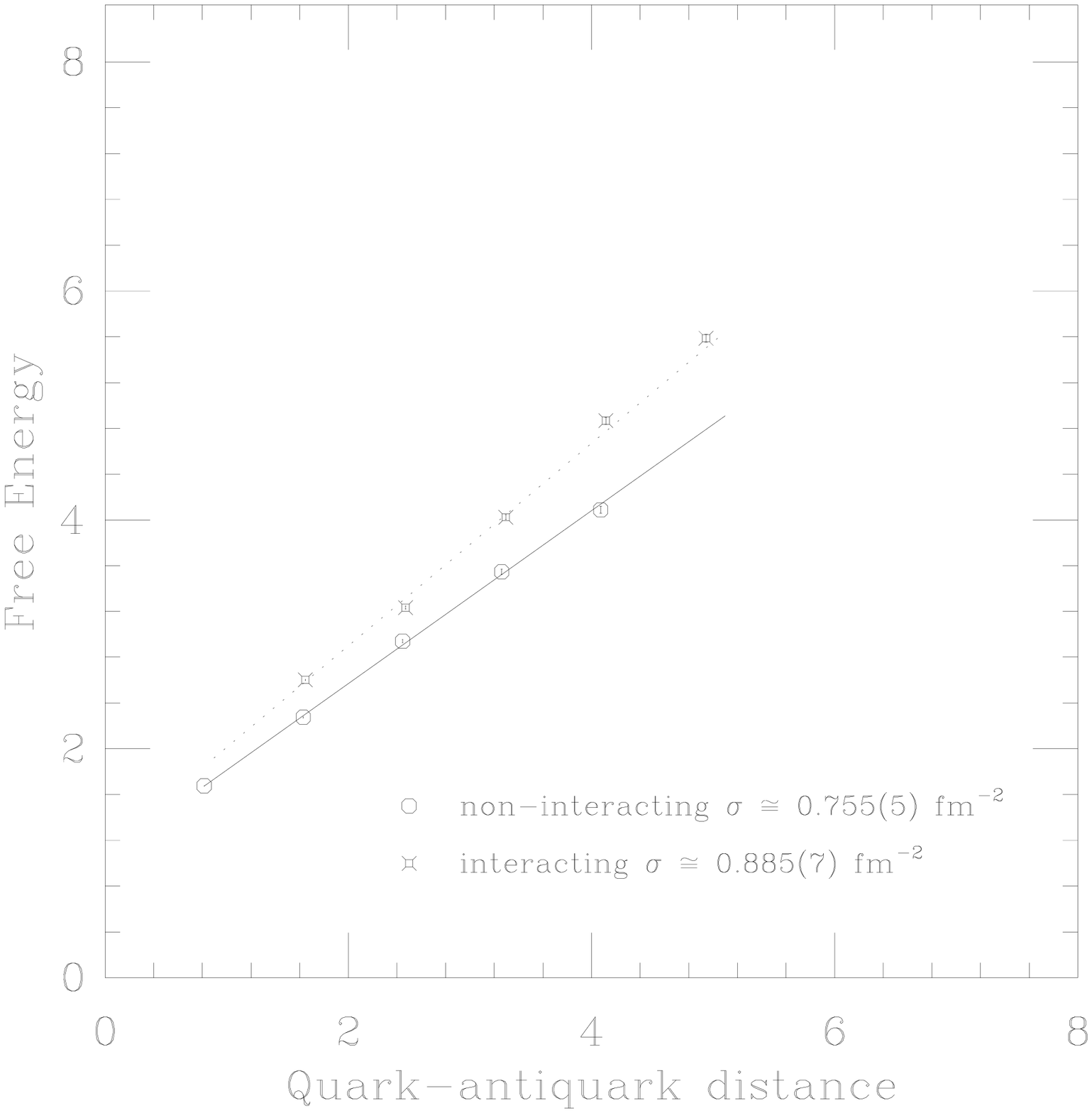}
  \caption{$LT=20$ }
  \label{fig:20}
\end{subfigure}%
\hspace{30pt}
\begin{subfigure}{.45\textwidth}
  \centering
  \includegraphics[width=1\linewidth,height=1\linewidth]{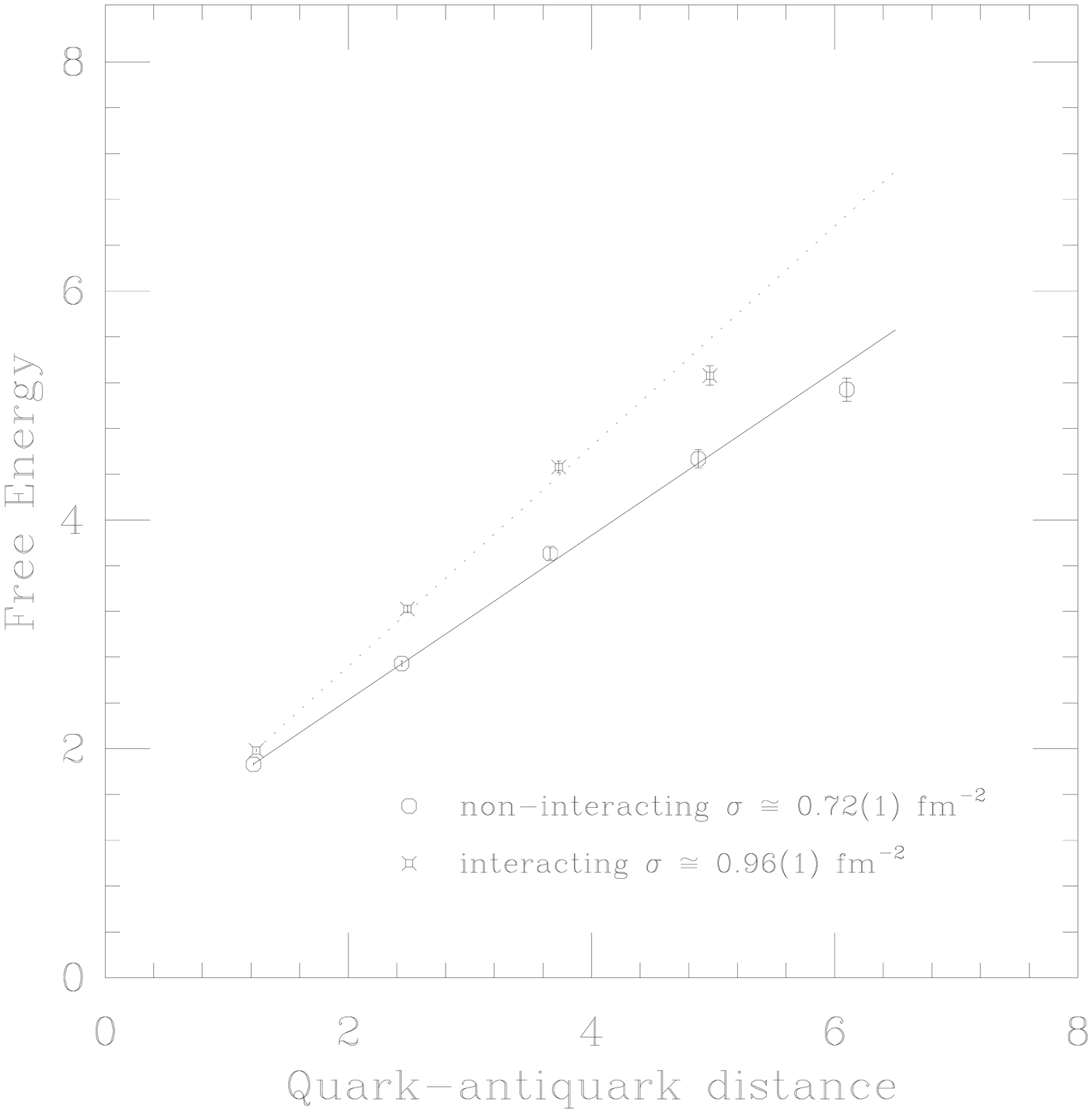}
  \caption{$LT=30$}
  \label{fig:30}
\end{subfigure}
\hspace{30pt}
\caption{The linear dependence of the free energy on the quark-antiquark separation for noninteracting and interacting dyon ensembles for $LT=20$ and 30. $\rho /T^{3}$ is fixed to one. The free energy grows linearly as the quark-antiquark separation increases. We are very close to the deconfinement temperature, $T=312 $ MeV.}
\label{fig:2030}
\end{figure}

For both noninteracting and interacting ensembles, the simulations are done for maximally nontrivial holonomy corresponding to the confinement phase, as described in Sec. \ref{sec:dyon}. Therefore, we expect that the potential grows linearly by increasing the quark-antiquark separation. As an example, Fig. \ref{fig:2030} illustrates this linear dependence for $LT = 20$ and 30 for noninteracting and interacting simulations.

\begin{table}[t]\footnotesize
\captionsetup{font=footnotesize}
 \begin{center}
    \begin{tabular}{| l | l | l | l |}
    \hline
    $ n_{D} $ & $ LT $ & configurations  \\ \hline
    1000 & 10 & 1600  \\ \hline
    8000 & 20 & 800 \\ \hline
    27000 & 30 & 120 \\ \hline
    125000 & 50 & 60 \\ \hline
    \end{tabular}
\end{center}
    \caption{Number of dyon configurations, number of dyons, $n_D$, and $LT$ for each simulation. $L^3$ indicates the spatial volume of the super cell and $T$ is the temperature.}
      \label{tab:input}
      \vspace{7ex}
\end{table}

To be able to compare the results of different simulations, we scale the data by the ansatz
\begin{equation}
 \frac{\sigma }{T^{2}} = \frac{\sigma(T=0)}{T_{c}^{2}}\left(\frac{T_{c}}{T}\right)^{2}A\left(1-\frac{T}{T_{c}}\right)^{0.63}\left(1+B\left(1-\frac{T}{T_{c}}\right)^{1/2}\right),
 \label{scale}
 \end{equation} 
where $B = 1 - 1/A$, $A=1.39$ \cite{5}, and $\sigma (T=0)=\left(440 MeV \right)^{2}$ corresponding to $T_{c}=312 $ MeV. Here, $\sigma$ indicates the string tension between the static quark and antiquark, and $T_c$  is the critical temperature. $\frac{\sigma }{T^{2}}$ (obtained from the plots like Fig. \ref{fig:2030}) is inserted into Eq. (\ref{scale}) and the corresponding temperature is obtained. Then, using the information in Table \ref{tab:input}, the lattice spacings are found for each simulation. As represented in Table \ref{tab:alldata}, the temperatures of our simulations are very close to the deconfinement temperature, $T=312 $ MeV, for both noninteracting and interacting simulations. The spatial lattice spacings and string tensions for each simulation are listed in Table \ref{tab:alldata}.
\\Since we use the interpolated original charges on the lattice, we should show that this approximation and the space discretization do not affect our results. In fact, we should show that the string tensions obtained from the lattices with different lattice spacings are equal at the same temperature. For both interacting and noninteracting ensembles, one can learn from Table \ref{tab:alldata} that the string tensions of the lattices with the same temperature agree very well within the errors. For example, for a noninteracting ensemble, for $LT=20$ and $30$ for which the temperatures are almost equal, the string tensions agree within the errors. 
Thus, our lattice spacings are small enough to not encounter discretization error.

\begin{table}
\captionsetup{font=footnotesize}
{\scriptsize
\begin{tabular}{l|l*{6}{c}r}
 &LT & $\sigma /T^{2}$ & T (MeV)& $\sigma (fm^{-2}) $ & \tiny{lattice spacing (fm)} & $\sigma (T)/\sigma (T=0)$ & $T/T_{c}$\\
\hline
\multicolumn{1}{ c | }{\multirow{5}{*}{\tiny{noninteracting}} } \\
& 10 & 0.46(1) & 295.31 & 1.01(1) & 0.44 & 0.20 & 0.946   \\ 
& 20 & 0.321(3) & 302.02 & 0.76(1) & 0.81 & 0.15 & 0.968   \\ 
& 30 & 0.304(7) & 302.80 & 0.72(1) & 1.21 & 0.14 & 0.970   \\ 
& 50 & 0.28(1) & 303.89  & 0.62(1) & 2.02 & 0.124 & 0.974   \\
\hline
\multicolumn{1}{ c| }{\multirow{5}{*}{\tiny{interacting}} } \\
& 10 & 0.633(4) & 285.95 & 1.333(6) & 0.43 & 0.27 &  0.92  \\ 
& 20 & 0.384(5) & 298.9 & 0.885(7) & 0.824  & 0.18 & 0.958   \\ 
& 30 & 0.423(8) & 297.024 & 0.96(1) & 1.24 & 0.19 &  0.952  \\  
& 50 & 0.34(1) & 301.08  & 0.79(2) & 2.045 & 0.16 &  0.965  \\
\end{tabular}
}
\caption{The numerical results of the simulations for different $LT $ for interacting and noninteracting ensembles.  The string tension between the quark-antiquark pair increases when the dyons interact with each other.}
\label{tab:alldata}  
\end{table}
\begin{figure}
\captionsetup{font=footnotesize}
  \begin{center}
    \includegraphics[width=.55\linewidth]{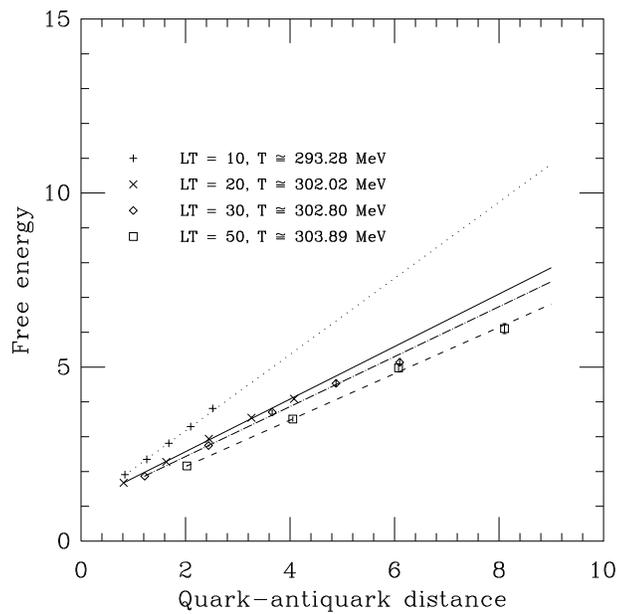}
    \caption{The scaled results of a noninteracting dyonic ensemble for different volumes.}
       \label{fig:sall}
  \end{center}
\end{figure}

\begin{figure}
\captionsetup{font=footnotesize}
  \begin{center}
    \includegraphics[width=.55\linewidth]{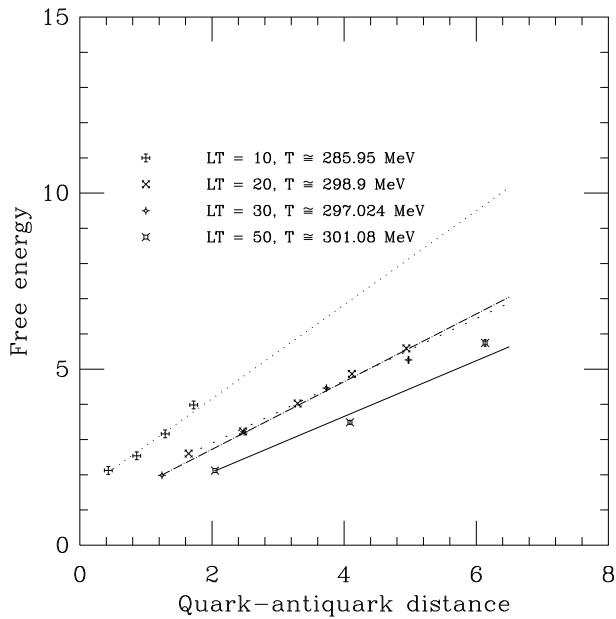}
    \caption{The same as Fig. \ref{fig:sall} but for the interacting dyonic ensemble.}
       \label{fig:isall}
  \end{center}
\end{figure}

\begin{figure}
\captionsetup{font=footnotesize}
  \begin{center}
    \includegraphics[width=.55\linewidth]{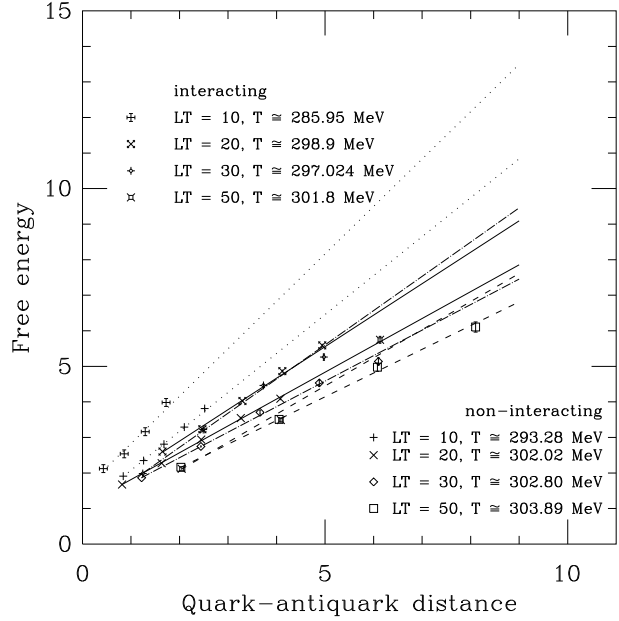}
    \caption{The scaled results of noninteracting and interacting simulations for different volumes. Comparison between the same values of $LT$ shows that when using interacting dyons, the string tension of the quark-antiquark pair increases.}
       \label{fig:bothdata}
  \end{center}
\end{figure}

\begin{figure}
\captionsetup{font=footnotesize}
  \begin{center}
    \includegraphics[width=.55\linewidth]{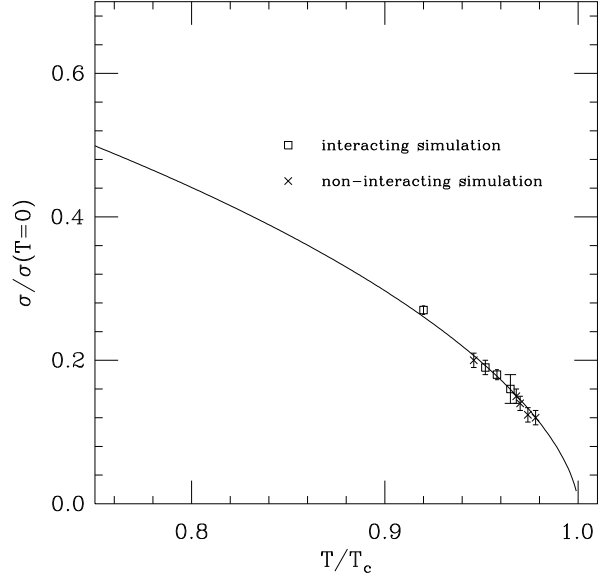}
    \caption{Our results fitted to the results of lattice gauge theory [Eq. (\ref{scale})]. }
    \label{fig:sigma}
  \end{center}
\end{figure}

Figures \ref{fig:sall} and \ref{fig:isall} illustrate the results of noninteracting and interacting simulations for different $LT$, after scaling. In general, as the temperature increases the string tension decreases, as one expects from the ansatz (\ref{scale}). To get the interacting results, we add dyonic interactions to the lattice of the noninteracting ensemble for the same $LT$. Therefore, we can compare the noninteracting and interacting results for each $LT$. As indicated in Table \ref{tab:alldata}, by adding the Coulombic interaction to the dyon ensemble the confinement temperature decreases slightly. The string tension of the quark-antiquark pair increases for the interacting ensemble.
This is a nice result. The interpretation is as follows: the interaction between dyons increases the free energy between the quark antiquark-pair, as the plots show. This means that the quark-antiquark pair system is more stable now and is further from the deconfinement phase compared with the noninteracting dyonic ensemble. In other words, it seems that the interaction between dyons increases the gluonic field strength compared with the noninteracting dyons. This explains the decrease in temperature for the same lattice when we just add the dyonic interaction to the noninteracting ensembles. Figure \ref{fig:bothdata} shows the results of interacting and noninteracting ensembles in one plot. Since the free energy is scaled, the slopes of the same "$LT$" simulations which show the string tensions between the static quark and antiquark can be compared easily between the interacting and noninteracting dyonic ensembles. A quantitative comparison is shown in Table \ref{tab:alldata}. Our simulation results are fitted to the plot of Eq. (\ref{scale}) in Fig. \ref{fig:sigma}.

For all noninteracting diagrams the order of interpolation $2p$ [in (\ref{exp}) of Sec. \ref{subsec:pme}] is equal to 4. This means that the charge of each dyon is interpolated to the four nearest neighbor points of the dyon location. But it seems that the $2p = 4$ is not enough for interacting simulations because of correlations between the dyon charges. Hence, we use $2p = 8$ for interacting dyons. We tried $2p=6$ and $2p=8$ for the noninteracting case and $2p=6$ for the interacting case and the results did not changed.

To show how good our choice $K_{i}=16$ is, we tried $K_{i}=8$ and $K_{i}=10$ for $LT=30$ for the noninteracting dyonic system. The errors on $\sigma/T^{2}$ are $21$\% and $8$\% for $K_{i}=8$ and $K_{i}=10$, respectively. Therefore, it seems that $K_{i}=16$ is a good choice. Increasing the parameter $K$ to the higher values does not give us a better estimation of the string tension, but the operating time increases drastically.  

By increasing the number of dyons, the effective charge becomes more efficient and a better result is expected. However, since we fix the parameter $\rho / T^{3} = 1$ in our simulations, the volume of the lattice would be increasing without increasing the number of lattice points, and therefore we get larger lattice spacings and larger errors. Therefore, there is a compromise between increasing the number of dyons and not getting a larger lattice spacing error. Table \ref{tab:alldata} shows that we are on the safe side.

As mentioned in Sec. \ref{sec:dyon}, adding antidyons changes the string tensions by a constant factor from physical results, $\sigma \rightarrow  \sqrt{2}\sigma$ \cite{1}. This affects the value of the temperature, although the system remains close to the deconfinement phase. However, our main results - the linearity of the free energy and the increase of the string tension due to the interaction - do not change.

\section{Conclusion}
\label{sec:Conclusions}
We have computed the free energy of a static quark-antiquark pair as a function of their separation by studying the Polyakov loop correlator for noninteracting and interacting dyon ensembles. We first applied the PME method to the dyons located randomly in different volumes to interpolate their charges on a 3D lattice with fixed dimensions, and then applied Ewald's method to this new system. As one expects, the free energy grows linearly as the separation increases. However, the string tension between the static quark-antiquark pair increases for the interacting dyonic ensemble. It seems that the dyonic interaction increases the gluonic strength, as expected.

\section*{Acknowledgement}

We are grateful to the research council of the University of Tehran for supporting this study.

\end{document}